\documentclass[prd,floats,preprint,superscriptaddress,eqsecnum,floatfix,nofootinbib,12pt,tightenlines]{revtex4}
\usepackage{amsmath}
\usepackage{graphics, graphicx}
\usepackage{epsfig}
\usepackage{ulem}
\usepackage{color}

\def\bh{\hat a}
\def\ch{\hat\phi}

\def\cF{{\cal F}}

\def\be{\begin{equation}}
\def\ee{\end{equation}}
\def\beq{\begin{eqnarray}}
\def\eeq{\end{eqnarray}}

\def\rf{{}}
\def\sf{}
\def\p0{\phi_0}
\def\z0{\zeta_0}

\def\aoi{D^{\ge 1}}

\def\pl{{pl}}
\def\tf{{}}
\def\uf{{}}
\def\vf{{}}
\def\wf{{}}
\def\yf{{}}
\def\zf{{}}
\def\af{{}}
\def\rf{{}}

\begin{document}

\title{Replication Regulates Volume Weighting\\in Quantum Cosmology}

\author{James  Hartle}
\affiliation{Department of Physics, University of California, Santa Barbara,  93106, USA}
\author{Thomas Hertog}
\affiliation{APC, UMR 7164 (CNRS, Universit\'e Paris 7), 10 rue A.Domon et L.Duquet, 75205 Paris, France\\ {\it and}\\
Intl Solvay Institutes, Boulevard du Triomphe, ULB -- C.P. 231, 1050 Brussels, Belgium}

\bibliographystyle{unsrt}

\begin{abstract} 
Probabilities for observations in cosmology are conditioned both on the {\uf universe's}  quantum state and on local data specifying the observational situation. We show the quantum state  defines a measure for prediction through such conditional probabilities that is well behaved for spatially large or infinite universes when the probabilities that our data is replicated are taken into account.
In histories where {\tf our}  data are rare volume weighting connects top-down probabilities conditioned on both the data and the quantum state to the bottom-up probabilities conditioned on the quantum state alone. We apply these {\yf principles}  to a calculation of the number of inflationary e-folds in a homogeneous, isotropic minisuperspace model with a single scalar field moving in a quadratic potential. We find that volume weighting is justified and  the top-down probabilities favor a large number of  e-folds.

\end{abstract}

\vspace{1cm}


\maketitle

\section{Introduction}
As observers we are physical systems within the universe. This paper develops the consequences of this elementary truth for the general nature of prediction in quantum cosmology. Specifically we provide a derivation  of a `measure' for prediction that is well-behaved for spatially very large or infinite universes.  The use of this measure is illustrated by a calculation of the predictions of Hawking's no-boundary quantum state (NBWF) \cite{Haw84} for the number of e-folds of inflation in a simple minisuperspace model. 

Quantum cosmological probabilities of use to us\footnote{Most generally these data would include a description of us as physical systems within the universe. It might be clearer to call the collection of human observers working on cosmology the human scientific IGUS (information gathering and utilizing system) as we have elsewhere \cite{HS07,Gel94}. But here we abbreviate this by `we', `us', etc.} are conditioned on some part of our data. For instance, the probability of an observation of the CMB spectrum {\wf is necessarily conditioned} on {\it when} and {\it where} the observation is made in the history of the universe.  Calculation of probabilities conditioned on our data must take account of the physical processes that produced us. {\uf There is a quantum probability that our data occur in any spacetime volume.  Therefore}, we do not necessarily exist in the universe, and, if we do, we are not necessarily unique. Indeed, in a very large universe the probability become significant that our data will be replicated {\wf exactly} elsewhere.  {\it Top-down} probabilities  conditioned on part of our data as well as the NBWF can differ significantly from the {\it bottom-up }probabilities conditioned\footnote{All probabilities in this paper are implicitly conditioned on the NBWF and the theory of dynamics, but we will not indicate this explicitly.} only on the NBWF. 

Top-down probabilities obtained by volume weighting of bottom-up probabilities have been discussed in the context of homogeneous, isotropic minisuperspace models by {\uf a number of authors}  \cite{Page97, Hawking07,HHH08a,HHH08b}. Consider the top-down probabilities conditioned on data on our past light cone that approximately locate us in some Hubble volume somewhere on a surface of homogeneity in spacetime. Assume that there is one and only one instance of our data on this surface. Then the top-down probabilities are proportional to the bottom-up probabilities multiplied by the number of Hubble volumes on the surface. A detailed derivation of this is given in Section \ref{prediction}, but roughly the top-down probabilities favor larger universes because there are more places for our data to be. 

Volume weighting evidently breaks down for universes with very large  or infinite spatial volume. However, that is {\wf also} the limit in which the probability for replication of our data becomes significant. For such histories a  more general weighting applies that depends on the probability $p_E$ that our data occur in any one Hubble volume. In Section \ref{prediction} we show explicitly that the  top-down probabilities that take account of the probability of replication $p_E$  provide a measure for prediction that remains well behaved even in the large or infinite volume limit. The resulting probabilities may depend significantly on $p_E$ but they are not divergent. Volume weighting is recovered for finite universes when $p_E$ is sufficiently small. 

Realistic values of $p_E$ will be very small but very, very difficult to compute precisely. However as we show in Section \ref{prediction},  $p_E$ only needs to be bounded to obtain results that are insensitive to its value by justifying volume weighting.  Such bounds are discussed in Section \ref{replication}. In Section \ref{inflation} we apply these bounds to derive volume weighting for the class of minisuperspace models considered in \cite{HHH08a,HHH08b}. {\zf This supports the conclusion of those papers where it was shown that top-down probabilities with volume weighting favor long periods of slow roll inflation.}

\section{Prediction in Quantum Cosmology}
\label{prediction}  

\subsection{Bottom-Up and Top-Down}
A quantum state of the universe predicts probabilities for the members of  decoherent sets of alternative coarse-grained histories of spacetime geometry and matter fields \cite{Hartle95,HHrules}.
An important example of {\uf such bottom-up probabilities}  is provided by the probabilities for the ensemble of possible classical histories of the  universe characterized by deterministic  correlations in time governed by the Einstein equation and the classical equations for matter fields \cite{HHH08a,HHH08b}. {\wf To achieve a discussion that is both managable and applicable we restrict attention in this paper to predictions of the properties of these classical histories, conditioned on data that is also part of their description.}

{As mentioned in the Introduction, useful predictions in cosmology  assume some part of our data  $D$  and predict conditional probabilities for other properties of the universe. These are called top-down probabilities \cite{Hawking06}.  The bottom-up NBWF probabilities for histories are inputs to the calculation of conditional probabilities. In general we take the point of view that all possible conditional probabilities are available in quantum cosmology. Which ones are useful to calculate is up to us. Two classes of top-down probabilities are of particular interest.

The first class {\tf consists of} probabilities for observations --- {\wf probabilities} for data that we seek to predict, either which we have now or might obtain in the future.  Probabilities for our observations of the universe are necessarily conditioned on data $D$ that include a local description of ourselves and our observational situation.  The use of top-down probabilities  {\vf to predict observations} is not a choice; they {\it are} the probabilities for {\it our} observations.

{\vf A second important class of top-down probabilities are {\uf those} for global properties of {\it our} universe  conditioned on our local data even when these global properties are not directly observable \cite{Hawking06}. Examples are the probabilities for past histories and for the nature of the structure of the universe on scales beyond the present horizon.} 

Much of our data $D$ result from chance accidents that have occurred over the history of the universe --- the chance accidents of biological evolution for instance.  {\vf The probability for this exact chain of accidents is very small in the observable spacetime volume.}   However, in a sufficiently large universe the probability becomes significant that even these accidents of biological evolution are repeated somewhere. For instance, in the oft considered model universe where many bubbles have nucleated with infinite volume spatial slices the probability is unity that our data occur an infinite number of times in {\it each} bubble for any non-zero $p_E$.  In a large universe it is both general and physically realistic to take account of the probability that the data $D$ may be  replicated elsewhere in the universe.  

{\vf When bottom-up probabilities are significant for multiple copies of our data at different locations in spacetime they do not specify which copy we are. To predict what we observe requires the specification of  a further (xerographic) distribution giving the probability that we are any particular copy \cite{HS09}. A simple and natural assumption is that we are equally likely to be any one of the copies. {\wf We will assume that here.} Further, to avoid venturing into the treacherous quagmire of current speculation concerning `Boltzmann brains' we will assume that we and the other copies are non-deluded ordinary observers.}

At best, our data  are limited to a spacetime region lying to the past of part of the past light cone of the present moment. All we know {\uf for certain} is that the universe exhibits {\it at least one instance} of a region with this data $D$ somewhere in {\uf classical} spacetime, a physical situation which we abbreviate as $\aoi$.   {\vf Assuming that we are equally likely to be any copy of our data, we 
calculate top-down probabilities conditioned on $\aoi$ by summing}  the bottom-up probabilities for entire classical histories weighted by the probability that  $D$ occurs at least once somewhere in spacetime.   In the following we  implement this idea concretely.

\subsection{Homogeneous and Isotropic Classical Ensembles}

In this subsection we follow an analysis in \cite{HS07} to give a general derivation of the weighting that connects top-down to bottom-up probabilities for the illustrative example of an ensemble of homogeneous, isotropic, classical,  Lorentzian {\wf cosmological histories}. We will apply this to the specific classical ensemble predicted by the NBWF in Section \ref{inflation}. But for the more general discussion here we need only assume that there is a one parameter family of such universes. We  denote the parameter by $\p0$ and the bottom-up probabilities by $p(\p0)$. We sketch the framework for constructing the top-down probabilities  for some feature $\cal F$ of the classical histories labeled by $\p0$ conditioned on one instance of a subset $D$ of our total data. The number of e-folds of scalar field driven inflation is the  feature treated in \cite{HHH08a,HHH08b} and in Section \ref{inflation} . 

In general there could be an instance of $D$ anywhere in a classical spacetime.  For example, if the spacetime  exhibits many nucleated bubbles with open spacelike slices inside, there could be an instance of $D$ in a large collection of bubbles at many different times.
But in homogeneous and isotropic {\vf models} it is reasonable to suppose that part of our data includes information about our location in time but not in space. That information could fix our location to be somewhere on one spacelike surface in a universe that starts from a singularity and expands forever. But there may be more than one spacelike surface on which our data could occur as in a bouncing homogeneous isotropic universe\footnote{{\af Looking beyond homogeneity and isotropy to spacetimes with nucleated bubbles there can be such surfaces in many different bubbles.}}. 

We therefore suppose  that $D$ can be divided into two parts: First, a part $D_s$ consisting of large scale observations that place the data on one or more sufaces of homogeneity $t_i(D_s,\phi_0)$ in each classical spacetime which we abbreviate simply by $t_i$. Observations of the present Hubble constant $H_0$  and local average energy density are an example. The second part $D_h$ consists of local observations  that are largely independent of the large scale features of the spacetimes. Observations of human observers, plants and animals, the features of the solar system, etc fall into this class along with a great many other details. Thus $D=(D_s,D_h)$.  For each $\phi_0$ divide the surface labeled by $t_i$  into Hubble volumes with size $\sim 1/H_0$ and denote their total number by $N_h(t_i,\phi_0)$.

Denote  by $p^i_E(D)$ the probability that the data $D$ occur in any {\sf  one of the Hubble volumes on the surfaces $t_i$ and assume that  the probability of more than one occurrence in any one volume is negligible. Although it is not necessary, for simplicity we will assume that $p^i_E(D)$ does not depend on global properties and in particular on the parameter $\p0$.  We will discuss these probabilities further in Section \ref{replication}.

As mentioned above, {\it all we know for certain from our local observations is that there is at least one occurrence of $D_h$ (abbreviated $\aoi_h$) in one of these Hubble volumes.} The probability that there is at least one instance of $D_h$ in the classical spacetime labeled by $\p0$  is $1$ minus the probability that there are no instances. {\af For any particular Hubble volume the probability that there is no instance of $D$ in it is $1-p^i_E(D)$. The probability that there is no instance of $D$ anywhere in the spacetime labeled by  $\p0$ is the product of such factors over all the $N_h(t_i,\p0)$ Hubble volumes  in a surface $t_i$ and then over all surfaces.} That is \cite{HS07} 
\begin{equation}
p(\aoi|D_s,\phi_0)= 1-\prod_i[1-p^i_E(D)]^{N_h(t_i,\phi_0)} 
\label{atleastone1}
\end{equation}
{\wf where here, as elsewhere, $t_i$ is understood to depend on $D_s$ and $\phi_0$.} 
Top-down probabilities are the bottom up probabilities $p(\phi_0)$ weighted by this probability, as we now derive.
{\af To avoid a debauche d'indices we will consider just the case where there is a single surface singled out by $D_s$ in all spacetimes in the ensemble. Then, dropping the now superfluous index $i$,} 
\begin{equation}
p(\aoi|D_s,\phi_0)= 1-[1-p_E(D)]^{N_h(t,\phi_0)}  
\label{atleastone2}
\end{equation}
{\af Results for more general cases {\rf will be discussed elsewhere}.

We construct the (top-down) probabilities $p({\cal F}|\aoi)$ for some feature $\cal F$ of the classical histories conditioned on at least one instance of a  subset $D$ of our total data. Denote by $\cal C_{\cF}$ the class of histories in the ensemble with the feature $\cF$. The probability $p(\cF|\aoi)$ is the sum of the probabilites for $\p0$ given $\aoi$ over all classical histories in this class, namely, 
\begin{equation}
p(\cF|\aoi) = \int_{\p0 \in {\cal C}_\cF }  d\p0 \ p(\p0|\aoi).
\label{pfeature}
\end{equation}
Introducing the characteristic function $e_\cF(\p0)$ for the class $\cal C_\cF$  and using the definition of conditional probability this can be written in terms of joint probabilities as\footnote{If there is a probability for $\cal F$ in a classical history $e_{\cal F}(\p0)$ can be replaced by $p({\cal F}|\phi_0)$.}
\begin{equation}
p(\cF|\aoi) = \frac{\int d\p0 e_\cF (\p0) p(\p0,\aoi)}{\int d\p0 p(\p0,\aoi)} \ .
\label{jntprobs}
\end{equation}
Now, 
\begin{equation}
p(\p0,\aoi) = p(\p0,D_s,\aoi_h) = p(\aoi_h|D_s,\p0) p(\p0,D_s) \ .
\label{twofour}\
\end{equation}
Further, 
\begin{equation}
p(\p0,D_s)=p(D_s|\p0)p(\p0) .
\label{twofive}
\end{equation}
Generally $p(D_s|\p0)$ will be constant over the range of $D_s$ for which there are surfaces contained in the history labeled by $\p0$ and zero otherwise.  

Combining \eqref{jntprobs}, \eqref{twofour}, \eqref{twofive}, and \eqref{atleastone1} we have%
\begin{equation}
p(\cF|\aoi) =\frac{\int d\p0 e_\cF (\p0) \{1-[1-p_E(D)]^{N_h(t,\phi_0)}\}p(D_s|\p0)p(\phi_0)}{\int d\phi_0 \{1-[1-p_E(D)]^{N_h(t,\phi_0)}\}p(D_s|\p0) p(\phi_0)} . 
\label{butotd}
\end{equation}
This central result can be summarized as follows: {\it To obtain the top-down probabilities for a feature $\cF$ of the classical histories conditioned on our data $D$, sum the bottom-up probabilities over those histories which contain  $\cF$ weighted by the probability \eqref{atleastone1} that there is at least one instance of the  data  $D$ somewhere in the universe.}

If the  data $D$ are dependent on anything like the chance accidents of biological evolution, the probabilities $p_E$ are well beyond our power to compute at the present. However, in certain important limits, the top-down probabilities become insensitive to the values of $p_E$. We describe two cases cases of this :

{\it The data $D$ are common:}  When the relevant values of  $N_h$ are very large compared to $1/p_E$, the data $D$ will be common in the universe.  In that limit the top-down probabilities are independent of $p_E$ and are sums over the bottom-up ones with no weighting.
\begin{equation} 
p(\cF|\aoi) \approx \int d\p0 e_\cF (\p0) p(\phi_0)  . 
\label{common}
\end{equation} 
This gives a well defined measure when the probabilities $p(\p0)$ are normalized, as we assume. This kind of limit {\rf plays a role} when the number $N_h$ {\rf can} become very large, as in some models of eternal inflation\cite{HHH09b}.

{\it The data $D_h$ are rare:} The data $D_h$ will be rare when there is a maximum number of Hubble volumes $N^m_h$ on any of the surfaces $t_i$ in {\tf each history with significant probability of} the classical ensemble and $p_E(D)\ll 1/N^m_h$. Then  \eqref{butotd} reduces to 
\begin{equation}
p(\cF|\aoi) \approx\frac{\int d\p0 e_\cF(\p0)  N_h(t,\phi_0)p(\phi_0)}{\int d\phi_0 N_h(t,\phi_0) p(\phi_0)} .
\label{lowpE}
\end{equation} 
This result is independent of $p_E$ and is exactly the volume weighting discussed in \cite{Page97,Hawking07,HHH08a,HHH08b}.
Volume weighting is {\wf thus} justified when {\vf there is a maximum number of Hubble volumes $N^m_h$ } and  $p_E(D)$ can be bounded by $1/N^m_h$. 

{\af Volume weighting of bottom-up probabilities becomes problematical if $N_h$ becomes very large or is infinite so that the integrals in \eqref{lowpE} diverge. That may be the case even for universes that have closed spatial slices if the surfaces determined by the data $D_s$
are infinite as in the interior of nucleated bubbles or for the reheating surface in certain inflationary models  
\cite{CREM}. However in ensembles of this kind the general expression \eqref{butotd} for top-down probabilities {\vf still} applies. This remains finite for large or even infinite $N_h$ even when the low $p_E(D)$ approximation \eqref{lowpE} to it breaks down. {\it The NBWF measure is finite when proper account is taken of the basic fact that we are physical systems within the universe that were formed by physical processes  that could also have occurred elsewhere. }}

\section{Objective Predictions}
\label{replication}

{\af All predictions of observations depend to some extent on where, when, and how the observations are made. The most useful predictions depend as little as possible on such details. They are then broadly applicable in many situations. Such predictions can be called objective. 

In the present models useful predictions are ones that depend as little as possible on the precise value of $p_E(D)$ and the data $D$ that determine it. We have identified two limits where this is the case: i) When the universe is so large that $D$ is  common, top-down probabilities are approximately equal to bottom up ones, and \eqref{common} holds. ii) When the universe is small enough that $D$ is rare and volume weighting described by \eqref{lowpE} applies. 

Data $D$ are rare on a surface specified by $D_s$ when $p_E(D) \ll 1/N_h^m$. To justify using the objective volume weighting limit it is only necessary to bound $p_E(D)$ from above. The data used to provide this bound could range from none of our present data to all of it.  The more data $D$ conditioned on, the smaller $p_E(D)$ will be. But the more data the more difficult it  will be to calculate $p_E(D)$ or even estimate it.   This suggests using more and more data $D$ for which $p_E(D)$ is estimable until it provides a bound that makes $D$ rare in the universe and volume weighting applicable --- if that is possible! 

This situation is not so different from that in every day experimental physics. Consider an experiment to measure the value of some constant. The results depend on the true value of the constant, but also to some small degree on the probabilities that the results are influenced  by details of the experimental arrangement.  The latter dependencies are the source of systematic errors. Systematic errors can be compensated for if their probabilities can be calculated accurately enough. If not, we seek to bound their probabilities from above thereby setting limits to the accuracy  of the measurement. Both compensation and bounds require a theory of the experimental arrangement.}


We can mention two strategies for identifying large amounts of data for which $p_E(D)$ is estimable:

{\it Naturally occuring data with a simple origin:}  Data on the observed temperature fluctuations in the CMB are an example. The CMB radiation originated from calculable small fluctuations in the early universe calculably propagated to the present. The probability of our CMB temperature map is something like $2^{-N_b}$ where $N_b$ is the number of bits necessary to describe the map --- a number of order $10^6$ for WMAP \cite{wmapteg}.
This is useful  $D$ for  predictions of the amount of past inflation but would not be appropriate for predictions of the CMB itself.

{\it Controllable random data:}  We can generate data under controlled circumstances whose probabilities are straightforward to compute. Commercially available quantum random number generators generate strings of random bits at the rate of several Mb/s. The probability $p_{\text{string}}$  of a string $N_b$ bits long is $2^{-N_b}$.  Run for a year such a generator will produce data whose probability is of rough order $p_{\text{string}} \sim 10^{-10^{13}}$. The probability $p_E(D)$ is this times the probability that there is at least one such machine in a Hubble volume. That may be difficult to estimate but $p_{\text{string}}$ provides a powerful upper bound. 

In the next section we will show the bounds provided by such kinds of data are sufficient to justify volume weighting for computing the probabilities for the amount of inflation in a homogeneous isotropic model quantum cosmology.

{\af The condition for the common limit \eqref{common} is $p_E(D) \gg 1/N_h^m$.  If $N^m_h$ is truly infinite this is trivially satisfied. But it $N_h^m$ is large but finite an estimate of $p_E(D)$ for {\it all} of our data $D$ is required. That may be difficult to compute or even define. 

 In classical physics it was possible to hope for a description of the universe that was independent of who observed it and how they did it.  Physics strives for such objectivity today but developments of the last century make it more difficult to achieve. In textbook quantum theory the description of a measured subsystem depends on what is measured. In cosmology observers and their apparatus are part of the universe not somehow separate from it. Predictions of global observations depend on where  and when they are made. In quantum cosmology an observer is a quantum subsystem like many others with a probability for evolving in any spacetime volume and a probability for being replicated elsewhere. 

In this paper we have sought to consider an observer as a quantum mechanical system within the universe although only in a very crude model.  We have not thereby abandoned the search for objective descriptions. Rather we have shown what is necessary to achieve them. }

\section{Probabilities for Inflation}
\label{inflation}
 
 In this section we illustrate the framework developed in Section \ref{prediction} by estimating as a function of $p_E$  the top-down probabilities predicted by the NBWF for the number of e-folds of scalar field driven inflation {\wf in the minisuperspace models considered in \cite{HHH08a,HHH08b}. These assume homogeneous, isotropic spacetime geometries and a single homogeneous scalar field $\phi$ moving in a quadratic potential  $V=(1/2)m^2\phi^2$.}  
The feature $\cal F$ in \eqref{butotd} is thus the number of e-folds $N$. The top-down probabilities conditioned on the NBWF and at least one instance  of a subset of our data $D$  are $p(N|\aoi)$. 

\subsection{Bottom-Up Probabilities for the Number of E-folds}
 
The bottom-up probabilities for the Lorentzian histories in the classical ensemble predicted by the NBWF were calculated in \cite{HHH08a,HHH08b}. We briefly review the essential results here {\vf specializing for simplicity to the case where the cosmological constant vanishes. }

In quantum cosmology states are represented by wave functions on the superspace of three-geometries and spatial matter field configurations. For homogeneous isotropic  models  minisuperspace is spanned by the scale factor $b$ and the value $\chi$ of the homogeneous scalar field. Thus, $\Psi = \Psi(b,\chi)$. 

The no-boundary wave function  (NBWF) \cite{Haw84} is defined by a sum-over-histories having the schematic form
\begin{equation}  
\Psi(b,\chi) =  \int_{\cal C} \delta g \delta \phi \exp(-I[a(\tau),\phi(\tau)]) .
\label{nbwf}
\end{equation}
Here, $a(\tau)$ and $\phi(\tau)$ are the histories of the scale factor and matter field and $I[a(\tau),\phi(\tau)]$ is their Euclidean action. The sum is over cosmological geometries that are regular on a manifold with only one boundary at which $a(\tau)$ and $\phi(\tau)$ take the values $b$ and $\chi$. The integration is carried out along a suitable complex contour ${\cal C}$ which ensures the convergence of \eqref{nbwf} and the reality of the result. We use units where $\hbar=c=G=1$.  

For some regions of minisuperspace the integral in \eqref{nbwf} can be approximated by the method of steepest descents. Then the wave function will be well approximated  to leading order in $\hbar$ by a sum of terms of the form 
\begin{equation}
\Psi(b,\chi) \approx  \exp[-I_R(b,\chi) +i S(b,\chi)] ,
\label{semiclass}
\end{equation}
one term for each extremizing history.  The functions $I_R(b,\chi)$ and $-S(b,\chi)$ are the real and the imaginary parts of the action evaluated at the extremum. 
In simple cases these extremizing histories may real;  but in general they  will be complex --- ``fuzzy instantons''.

 A wave function of the semiclassical (WKB) form \eqref{semiclass} predicts an ensemble of coarse-grained Lorentzian histories $(\bh(t),\ch(t))$ in regions of minisuperspace where  $S(b,\chi)$ varies sufficiently rapidly when compared with  $I_R(b,\chi)$. {\wf The requirements for this are called the `classicality conditions'. When they are satisfied,} the histories are the integral curves of $S(b,\chi)$.  Their probabilities to leading semiclassical order are given by $\exp[-2I_R(b,\chi]$. This is constant along the integral curves as a consequence of the Wheeler-DeWitt equation.  
 
 There is a one-parameter family of extremizing histories that can be labeled by the magnitude of the complex scalar field $\phi_0\equiv |\phi(0)|$ at the `South Pole' of the fuzzy instanton.  {\vf The NBWF thus predicts a one-parameter ensemble of}  classical Lorentzian solutions conveniently also labeled by $\phi_0$.  The classicality  condition is satisfied for $\phi_0$ greater than a critical value $\phi_0^c \approx 1.2$.} The bottom-up probabilities for classical Lorentzian histories can therefore be written in leading semiclassical order as
 \begin{equation}
 p(\phi_0) \approx \exp[-2I_R(\phi_0)]   \quad\quad (\phi_0>\phi_0^c)  
 \label{buprobs}
 \end{equation}
{\wf  and are zero in this semiclassical approximation for $\phi_0<\phi_0^c$.}
 The results for a numerical calculation for $I_R(\phi_0)$ are shown in the left panel of Figure \ref{inflationfig}.
\begin{figure}[t]
\includegraphics[width=3.2in]{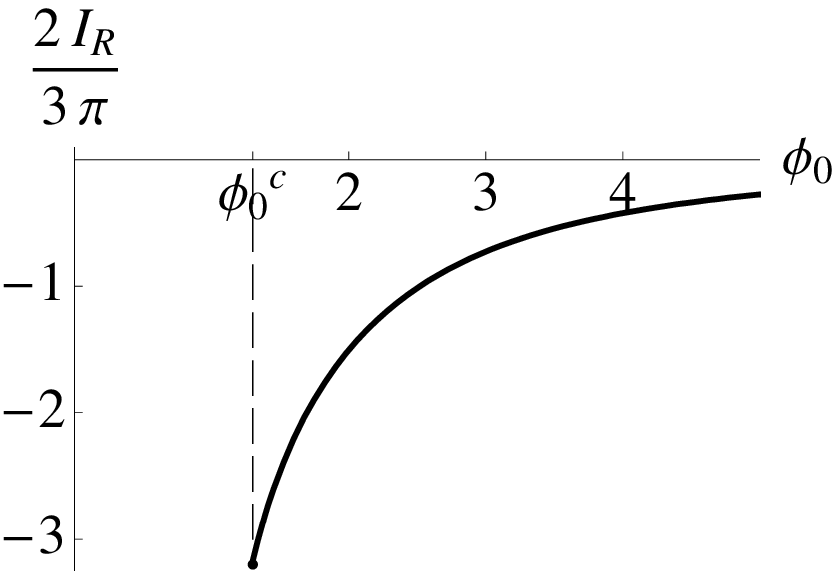} \hfill
\includegraphics[width=3.2in]{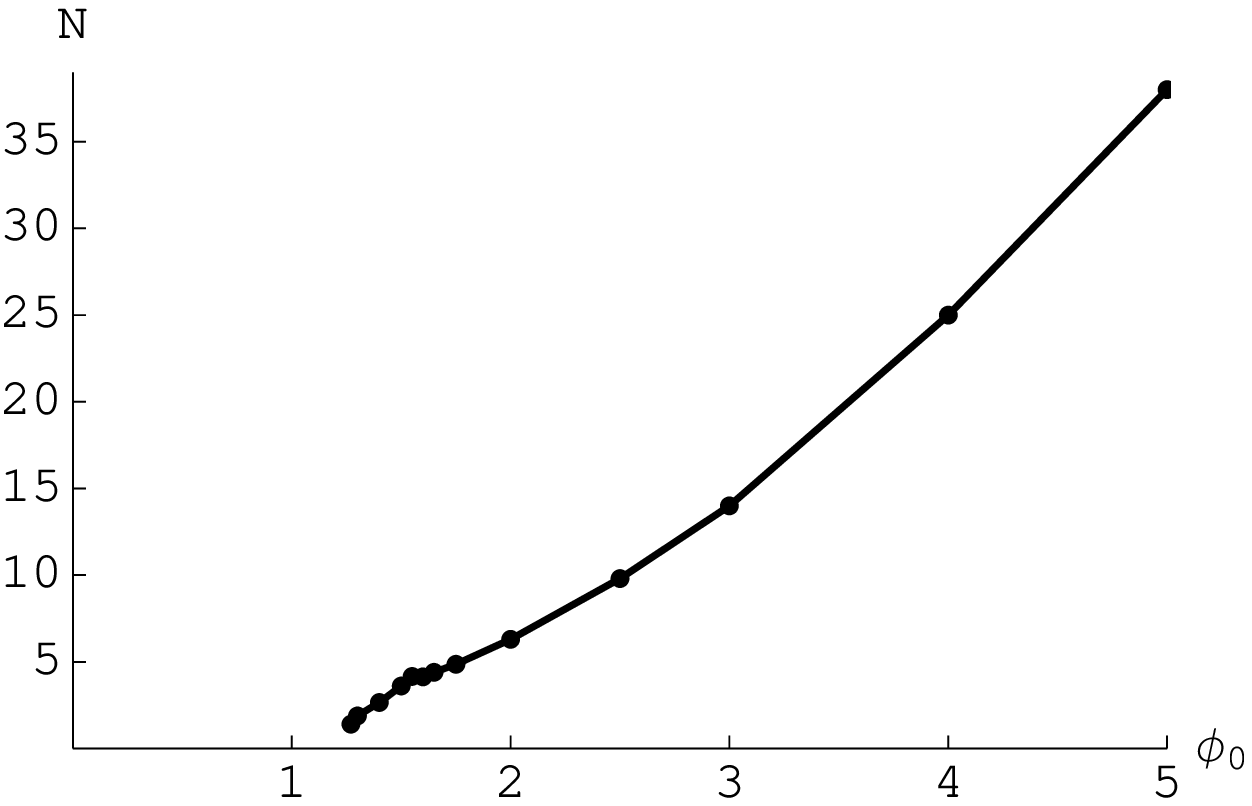} 
\caption{{\it Left panel}: The different histories in the ensemble predicted by the NBWF can be labeled by $\phi_0$  --- the magnitude of the scalar field at the South Pole of the fuzzy instanton.  The asymptotic value of the real part of the action of the complex solutions that behave classically at late times is plotted here as a function of $\phi_0$. The action tends to a finite value at the lower bound $\phi_0^c$ that arises from the classicality conditions. It goes to zero as $\sim -\pi/ 2 (m \phi_0)^2$ at large $\phi_0$. The classical ensemble ranges from $\phi_0^c$ to $\p0^\pl$ at the Planck scale.  The results shown here are for $m^2=.05$ {\vf (Planck units)}.\\ 
{\it Right panel}: The number of e-folds $N$ of inflation in the different classical histories predicted by the NBWF. {\wf Without further constraints the NBWF} selects inflating universes but  the bottom-up probabilities favor histories with a small number of e-folds.}
\label{inflationfig}
\end{figure}

A striking feature of the  ensemble of classical histories in this model is the close connection between classicality and inflation \cite{HHH08b}. The histories have values of $\hat h(t)\equiv(d\bh/dt)/{\bh}$ and $\ch(t)$, which {\it all} lie within a very narrow band around $\hat h = m \ch$ characteristic of Lorentzian slow roll inflationary solutions. Since the histories represented in Fig {\ref{inflationfig}  {\vf span} the full ensemble predicted by the NBWF, it follows that a classical homogeneous and isotropic universe {\it must have}  an early inflationary state if the universe is in the no-boundary state. This is remarkable since inflationary spacetimes encompass an extremely small subset in classical phase space \cite{Gibbons06}.

However, as Figure \ref{inflationfig} shows, the bottom-up probabilities conditioned only on the NBWF are largest for classical histories with a small amount of inflation. We next estimate the probabilities for the number of e-folds in our universe defined by the top-down probabilities conditioned on data {\uf that, among many other things, specify} our location in time. 
 
 \subsection{Top-Down Probabilities for the Number of E-folds} 
 
 For a discussion of the number of inflationary e-folds the relationship between top-down and bottom-up probabilities at large $\phi_0$ is of special interest. That is where the number of e-folds is the largest (Figure \ref{inflationfig}, right). 
 
For sufficiently large $\p0$ there is an approximate analytic solution for the fuzzy instanton giving the NBWF in the semiclassical approximation \cite{Lyo92}. The solution is the complex analog of the familiar `slow roll' approximation for motion in a potential $(1/2)m^2\phi^2$. 
The predictions for the ensemble of classical histories in this approximation were derived in \cite{HHH08a,HHH08b}. We next quote the results relevant for this discussion\footnote{{\tf The second line of eq.(6.1) in \cite{HHH08b} should read $\hat a (t) = a(y(t)) \sim e^{\mu \phi (t) t +\mu^2 t^2/6}$. The solution for the scale factor in a model with zero cosmological constant such as we consider here is obtained by replacing the parameter $\mu$ in this formula by $m$} {\vf in this case where $\Lambda=0$.}}.

The predicted classical ensemble consists of Lorentzian histories of the form: 
\begin{subequations}
\label{soln}
\begin{align}
\ch(t) &\approx \phi_0 - mt/3    \label{chisoln} , \\
\bh(t) &\approx  \frac{1}{m\p0} \exp[mt(\p0-mt/6)]  \label{bsoln} 
\end{align}
\end{subequations}
assuming that $t$ is not so large that the slow roll approximation for the fuzzy instanton fails. These are Lorentzian, slow roll, inflationary solutions to the Einstein equation with the scalar field approximately $\p0$ at the start of inflation.

Denote by $N(\p0)$ the number of inflationary e-folds in the classical history labeled by $\p0$.  This is
\begin{equation} 
N(\p0) \equiv \int_0^{t_e} dt  \frac{1}{\bh(t)}\frac{d\bh}{dt} \approx \frac{3}{2}\p0^2 .
\label{nefolds}
\end{equation}
The integral is from the start of inflation at $t=0$ with $\ch\approx\p0$  to its end at $t_e$ with $\ch=\phi_e$.  The approximation assumes that $\p0\gg \phi_e \approx 1 $. 

The  real part of the action of the fuzzy instanton in this approximation is
\begin{equation}
I_R (\p0)\approx -\frac{\pi}{2(m\p0)^2}.
\label{action}
\end{equation}
This determines the bottom-up probabilities of the histories $p(\p0)$  through \eqref{buprobs},
\begin{equation}
p(\p0)\approx \exp[\pi/(m\p0)^2]\approx \exp[2\pi/(3m^2N)] .
\label{buprobs1}
\end{equation}

Eqs \eqref{buprobs1} and \eqref{nefolds} provide the ingredients necessary to estimate the conditional (top-down) probability $p(N|\aoi)$ for the number of e-folds of our universe given at least one instance our data $D$. Suppose for simplicity that  our data locate us on a unique surface of homogeneity in each spacetime of the classical ensemble. The {\vf one to one} relationship between $N$ and $\p0$ provided by \eqref{nefolds}  [cf. Fig \ref{inflationfig}] allows $N$ to be used as a label for histories. The number of present Hubble volumes in a history with $N$ e-folds will be
\begin{equation}
N_h(N) = N^0_h(N) \exp(3N)  
\label{nh}
\end{equation}
where $N^0_h(N)$ varies slowly with $N$ and depends on the present Hubble constant. 

The resulting estimate of \eqref{butotd} is the following
\begin{equation}
p(N|\aoi) \approx \frac{\{1-[1-p_E(D)]^{N^0_h(N)\exp(3N)}\}\exp[3\pi/2m^2 N]/N^{1/2}}{\int (dN/N^{1/2}) \{1-[1-p_E(D)]^{N^0_h(N)\exp(3N)}\}\exp[3\pi/2m^2 N]}. 
\label{efold}
\end{equation}
where $p_E (D)$ is the probability that the data $D$ occur in any one of the Hubble volumes {\yf and we have used \eqref{nefolds} to convert the measures. }

The range of $N$ is bounded below by the classicality constraint, which implies that $N \gtrsim 3/2$. We assume the range is bounded above by $N \lesssim N^{\pl} = 3/2m^2$  so that  the energy density in the scalar field is less than the Planck density. The normalizing integral in the denominator therefore converges\footnote{In models where the potential becomes flat at large $\phi$, the range of $\phi_0$ may extend all the way to infinity. In this case one must include the prefactor for the no-boundary probability distribution to be normalizable \cite{barvinsky}.}.
\begin{figure}[t]
\includegraphics[width=5.0in]{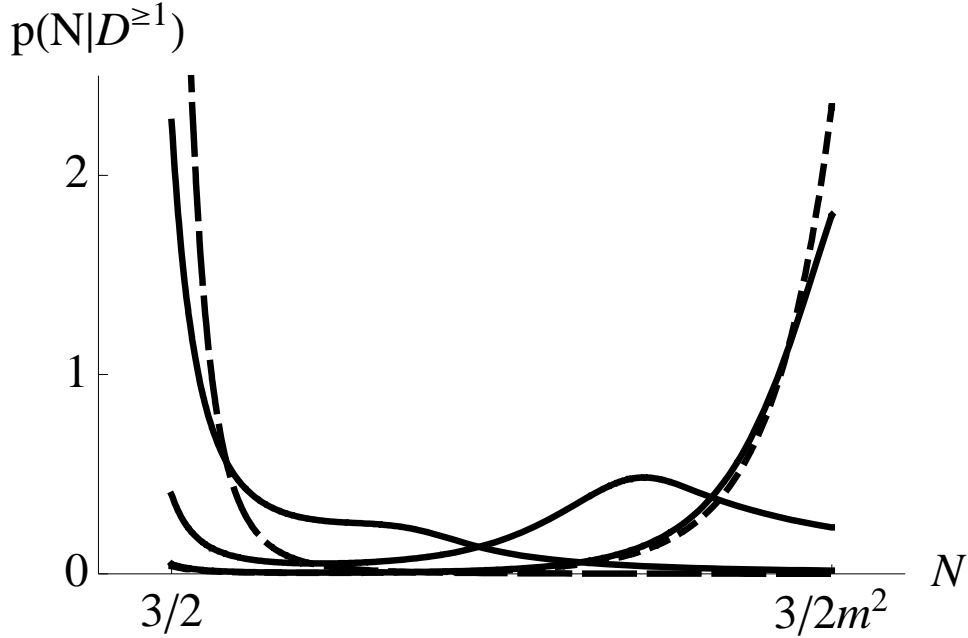} 
\caption{The top-down probabilities for the number of e-folds $N$ predicted by the NBWF conditioned on at least one instance of a subset of our data $D$ for five different values of $p_E (D)$.   The range of $N$ in this model is bounded from below by the classicality constraint which implies $N \gtrsim 3/2$. We assume it is also bounded above  by the Planck scale which means $N \lesssim N^{\pl} = 3/(2m^2)$. 
The dashed curve that rises to the left has $p_E=1$ and hence equals the bottom-up distribution \eqref{buprobs1}.
The dashed curve rising to the right, at large $N$, has $p_E \ll 1/N^{pl}_h$, and gives the volume weighted no-boundary probabilities. The three remaining curves correspond to (from top to bottom at large $N$) 
$p_E = 1/N^{pl}_h,\ 100/N^{pl}_h$ and $1/(N^{pl}_h)^{1/2}$.}
\label{TDefoldings}
\end{figure}

The top-down probabilities $p(N|\aoi)$ are shown in Figure \ref{TDefoldings}, for five different values of $p_E$. These range from $p_E =1$, for which the top-down weighting has no effect,  to $p_E \ll 1/N_h^{\pl}\sim \exp((9/(2m^2))$ for which the top-down probabilities are given by volume weighting to a good approximation. Figure \ref{TDefoldings} shows that top-down probabilities differ significantly from the bottom up ones for small $p_E$.  For realistic values of $m$ volume weighting holds for  $p_E \ll \exp(-10^{12})$. We have seen in Section \ref{replication} that at least in the homogeneous isotropic ensemble discussed here, one can easily find data $D$  for which $p_E$ meets this condition. Thus the volume weighting used in  \cite{Hawking07,HHH08a,HHH08b} is derived. And thus the prediction of those papers for a high probability that our universe underwent a significant amount of inflation in the past is justified in the context of our simple minisuperspace models.

\section{Conclusion}

The quantum state of the universe and the theory of its quantum dynamics\footnote{Supplemented by a xerographic distribution when necessary as discussed in the Introduction.}  are in principle adequate to predict probabilities for every physically meaningful set of alternatives the universe may exhibit. 
We have derived a general connection between two important sets of probabilities {\rf in quantum cosmology}. First, there is the set of  bottom-up probabilities for the alternative classical histories of the universe conditioned on the theory of the {\rf quantum} state and dynamics alone. Second is the set of top-down probabilities for the classical properties of our universe --- our observations, our history, etc. --- that is further conditioned on data that localize us to one or more spacelike surfaces in four-dimensional classical spacetime. 

The top-down probabilities \eqref{butotd} are appropriately weighted sums of bottom-up probabilities. 
This weighting is not a choice, or a postulate, or a proposal. Instead it arises necessarily {\wf within the usual framework of quantum mechanics  from just three}  considerations:
1) We, together  with our data, arose from quantum processes within the universe. We occur in any Hubble volume with a probability $p_E$ that is approximately independent of global features of the universe we seek to predict. 
2) In a large universe our data may be replicated  elsewhere with significant probability. But all we know for certain about this data is that the universe exhibits at least one instance of it. 3) {\wf We are equally likely to be any of the instances of our data that the universe exhibits.} 

Volume weighting arises as an approximation to this more general weighting when our data are rare in all histories in the ensemble that are predicted with any significant probability. Unlike its approximation, the general weighting \eqref{butotd} is well behaved even when spatial volumes become infinite. 

{\yf In \cite{HHH08a,HHH08b} it was explicitly assumed that our data  are rare in the universe. This paper has provided a quantitative justification of that assumption in terms of the probability that our data are replicated. That justification supports the conclusion, of both those papers and this, that top-down probabilities derived from the NBWF favor many efolds of slow roll inflation in simple minisuperspace models.}


It has not escaped our notice that the discussion in this paper may bear on issues that arise in eternal inflation.
We have seen that the NBWF provides a `measure' for {\rf global} predictions in cosmology that remains well defined even  when the ensemble of histories includes universes in which our data {\uf locate us on} one or more spatially infinite surfaces. Such universes occur in the regime of eternal inflation. Indeed it has been argued that, in the model we have discussed, the reheating surface can have infinite volume when one includes the effect of inhomogeneities \cite{CREM}. In a forthcoming paper \cite{HHH09b} we consider inhomogeneities explicitly and show how the NBWF measure of these can be applied to predict the structure of our universe on observable scales, as seen by a typical observer, in the regime of eternal inflation.

\acknowledgments  {\uf We thank Stephen Hawking for inspiration and guidance in quantum cosmology over a long period of time as well as discussions of this work. }We thank Matt Kleban, Mark Srednicki, Paul Steinhardt, Neil Turok and the participants of the Cosmic Singularity Symposium at PCTS (Princeton) for stimulating discussions. We also thank A. Barvinsky and M. Tegmark for discussions on particular points. We thank the Mitchell Institute of Texas A\&M University for hospitality while part of this work was being completed. The work of JH was supported in part by the National Science Foundation under grant PHY05-55669.

\nopagebreak


\begin{thebibliography}{99}

\bibitem{Haw84} S.W.~Hawking, {\it The Quantum State of the Universe},
{\sl Nucl.~Phys.~B}, {\bf 239}, 257-276 (1984).


\bibitem{Gel94} M.~Gell-Mann, {\sl The Quark and the Jaguar}, W.~Freeman
San Francisco (1994).

\bibitem{HS07} J.B. Hartle and M. Srednicki, {\it Are We Typical?}, {\sl Phys. Rev. D} {\bf 75}, 123523 (2007), arXiv:0704.2630. 

\bibitem{Page97} D. Page, {\it Space for both No-Boundary and Tunneling Quantum States of the Universe},  {\sl Phys. Rev. D}, {\bf 56}, 2065 (1997). 

\bibitem{Hawking07}
S.W.~Hawking, {\it Volume Weighting in the No Boundary Proposal},  arXiv:0710.2029. 

\bibitem{HHH08a} J.B.~Hartle, S.W. Hawking, and T. Hertog, {\it The No-Boundary Measure of the Universe}, {\sl Phys. Rev. Lett.},  {\bf 100}, 202301 (2008), arXiv:0711:4630.

\bibitem{HHH08b} J.B.~Hartle, S.W. Hawking, and T. Hertog, {\it Classical Universes of the No-Boundary Quantum State}, {\sl Phys. Rev. D} {\bf  77}, 123537 (2008), arXiv:0803:1663.

\bibitem{HHrules} J.B. Hartle and T. Hertog, {\it Classical Prediction in Quantum Cosmology},
in preparation. 

\bibitem{Hartle95} J.B.~Hartle, {\it Spacetime Quantum Mechanics and the Quantum Mechanics of
Spacetime}, in {\sl Gravitation and Quantizations: Proceedings of
the 1992 Les Houches Summer School}, ed.~by B.~Julia \& J.Zinn-Justin, North Holland, Amsterdam
(1995); arXiv:gr-qc/9304006. 

\bibitem{Hawking06}
S.W. Hawking and T. Hertog, {\it Populating the Landscape: A Top Down Approach},
{\sl Phys. Rev. D} {\bf 73}, 123527 (2006), arXiv:hep-th/0602091.


\bibitem{HS09} M. Srednicki and J.B.~Hartle, {\it Science in a Large Universe}, arXiv:0906.0042. 

\bibitem{HHH09b}  J.B. Hartle, S.W. Hawking, and T. Hertog, {\it The No-Boundary Measure in the Regime of Eternal Inflation}, in preparation.

\bibitem{CREM} S. Winitzki, {\it The Eternal Fractal in the Universe}, {\sl Phys. Rev. D} {\bf 65}, 083506 (2002), arXiv: gr-qc/0111048; 
P. Creminelli, S. Dubovsky, A. Nicolis, L. Senatore, M. Zaldarriaga, {\it The Phase Transition to Slow-Roll Eternal Inflation}, {\sl JHEP}, {\bf 0809}, 036 (2008), arXiv:0802.1067

\bibitem{wmapteg}  M. Tegmark, private communication. 

\bibitem{Gibbons06} G. W. Gibbons, N. Turok,  {\it The Measure Problem in Cosmology}, {\sl Phys. Rev. D}, {\bf 77}, 063516 (2008), arXiv:hep-th/0609095.

\bibitem{Lyo92} G.W.~Lyons, {\it Complex Solutions for the Scalar Field Model of the Universe}, {\sl Phys. Rev. D}, {\bf 46}, 1546-1550 (1992).

\bibitem{barvinsky} A.O. Barvinsky and A.Yu. Kamenshchik, {\it Quantum scale of Inflation and Partcle Physics of the Early Universe}, {\sl Phys. Lett. B}, {\bf 332},  270 (1994), arXiv: gr-qc/9404062.

\end{thebibliography}
\end{document}